\documentclass[letterpaper]{article}

\usepackage[T1]{fontenc}

\usepackage{geometry}
\usepackage{setspace}

\usepackage[style = chem-acs]{biblatex}
\usepackage{graphicx}
\usepackage{float}
\newfloat{scheme}{htbp}{los}
\floatname{scheme}{Scheme}
\floatname{chart}{Chart}
\newfloat{graph}{htbp}{loh}
\usepackage{caption}

\usepackage{chemformula} % Formulas using \ch{}
\usepackage{chemfig}
\usepackage[version = 4]{mhchem} % Formulas using \ce{}

\usepackage{authblk}
\author[1]{Line Mouaffac}
\author[1]{Maïwenn Souetre}
\author[1,2]{Guillaume Jeanmairet}
\author[1,2,3]{Mathieu Salanne}
\affil[1]{Sorbonne Université, CNRS, Physico-Chimie des Electrolytes et Nanosystèmes Interfaciaux, PHENIX, F-75005 Paris, France}
\affil[2]{Réseau sur le Stockage Electrochimique de l’Energie (RS2E), FR CNRS 3459, 80039 Amiens Cedex, France}
\affil[3]{Institut Universitaire de France (IUF), Paris 75231,
France}

\title{Competing Ring-Opening and Hofmann Elimination Pathways in Aqueous TEMPO Catholytes: A First-Principles Study}
\begin{document}

\maketitle

\begin{abstract}
Aqueous redox-flow batteries based on TEMPO derivatives are promising for large-scale energy storage, but their practical use is limited by the chemical instability of the oxidized \(N\)-oxoammonium state. In this work, we investigate the degradation of three TEMPO derivatives using \textit{ab initio} molecular dynamics combined with enhanced sampling. Two proposed degradation mechanisms, ring opening and Hofmann elimination, are examined and their corresponding activation free energies are compared. For all derivatives considered, ring opening exhibits a lower activation free energy than Hofmann elimination, identifying it as the kinetically preferred degradation pathway. The magnitude of the ring-opening barrier, however, varies significantly between molecules, showing that different functionalizations strongly influence its stability toward degradation. The predicted preference for ring opening is consistent with available experimental studies, which have identified or inferred ring-opening degradation for several TEMPO-based catholytes. These results provide an atomistic picture of degradation pathways that are difficult to resolve experimentally and highlight the importance of molecular structure in controlling the kinetic stability of TEMPO derivatives in aqueous electrolytes.
\end{abstract}

\section*{Keywords}
redox-flow batteries, TEMPO, degradation mechanism, \textit{ab initio} molecular dynamics, enhanced sampling, aqueous electrolyte, electrochemical stability

\section{Introduction}
As the global population continues to increase, there is a growing need for safe, environmentally-friendly, and low-cost battery storage systems. Historically, lead-ion batteries were among the first to be used for energy storage~\cite{armand2008a}, providing short bursts of power and small-scale storage. However, for mass energy storage (e.g., storage of solar or wind energy for the electrical grid), they have several drawbacks: short cycle life, low efficiency, and significant environmental risks. 
This is where redox flow batteries (RFBs) come into play. Unlike typical solid-electrode batteries, RFBs store energy in liquid electrolytes present in external tanks~\cite{noack2015chemistry}. These electrolytes are pumped through an electrochemical cell, where reduction and oxidation reactions (hence the term "redox") convert chemical energy to electricity and vice versa. This configuration allows for the power and energy to be decoupled: power depends on the size of the cell stack, while energy depends on the volume of the electrolyte. This makes RFBs highly scalable and particularly well-suited for stationary large-scale energy storage~\cite{zhang2018progress}. Since the electroactive species are dissolved in the electrolyte, the latter should exhibit several desirable characteristics such as high solubility and ionic conductivity to ensure the efficiency of the RFB~\cite{arevalo2021redox,fang2025degradation}. Vanadium-based RFBs are currently the most mature and widely deployed technology, offering advantages such as high reversibility and minimized cross-contamination due to the use of the same element in both half-cells~\cite{skyllas2011progress,lourenssen2019vanadium}. However, vanadium RFBs suffer from several limitations, including relatively low cell voltage constrained by the aqueous stability window, limited energy density, high materials cost, and exhibit performance loss due to thermal instability and precipitation of the redox species~\cite{cunha2015vanadium,dmello2016cost}. 

Organic molecules represent a promising alternative to address these limitations, as they offer structural tunability, lower cost per unit mass, and reduced environmental impact~\cite{luo2019status,winsberg2017redox,shoaib2024advances}. Among them are nitroxide radicals such as (2,2,6,6-tetramethylpiperidin-1-yl)oxyl (TEMPO) that have emerged as particularly attractive candidates for the positive electrolyte (catholyte), owing to their fast and reversible one-electron oxidation kinetics, high solubility in a variety of solvents, and chemical stability under cycling conditions~\cite{pedraza2023unprecedented,zhou2020fundamental}. The redox reaction involves the conversion between the nitroxide radical and its corresponding \(N\)-oxoammonium cation. However, the latter exhibits reduced stability as it is susceptible to nucleophilic attack by water molecules and anions present in the electrolyte~\cite{seo2023covalent,nolte2022stability}. In addition to this susceptibility, several degradation mechanisms have been proposed, including ring-opening degradation, Hofmann-like elimination, and chloromethane elimination~\cite{rohland2022structural,yue2026towards,tang2025adjusting,boutamine20253}. However, the degradation pathways remain hypothetical due to the lack of experimental characterization . This limitation arises primarily from the experimental difficulty to detect short-lived reactive intermediates of the \(N\)-oxoammonium cation. Moreover, chemical functionalization can impact the preferential degradation pathway in a subtle way that is difficult to guess \textit{a priori}. Consequently, \textit{ab initio} molecular dynamics (AIMD) emerges as a powerful tool to investigate such processes, as it naturally captures bond breaking and formation at the electronic structure level. AIMD has already been applied to TEMPO derivatives to investigate its redox properties in acetonitrile~\cite{reeves2020first}. Complementary classical molecular dynamics studies have also been used to characterize the solvation and local environment of TEMPO-based redox species~\cite{berthin2021solvation,goloviznina2023electrochemical}. Building on these studies, AIMD can be combined with enhanced sampling techniques to investigate chemical degradation pathways and quantify their associated free-energy barriers.
%However, AIMD is severely limited by accessible timescales, making it unable to sample rare events such as degradation reactions that occur on timescales far beyond those reachable by standard simulations. Enhanced sampling techniques, such as metadynamics~\cite{laio2002escaping,barducci2008well} and umbrella sampling (US)~\cite{torrie1977nonphysical}, bridge this gap by accelerating the exploration of free energy landscapes, enabling the study of reaction mechanisms and the quantification of their associated energy barriers~\cite{henin2022enhanced}. \\

\begin{figure}[!h]
    \centering
    \includegraphics[width=0.5\linewidth]{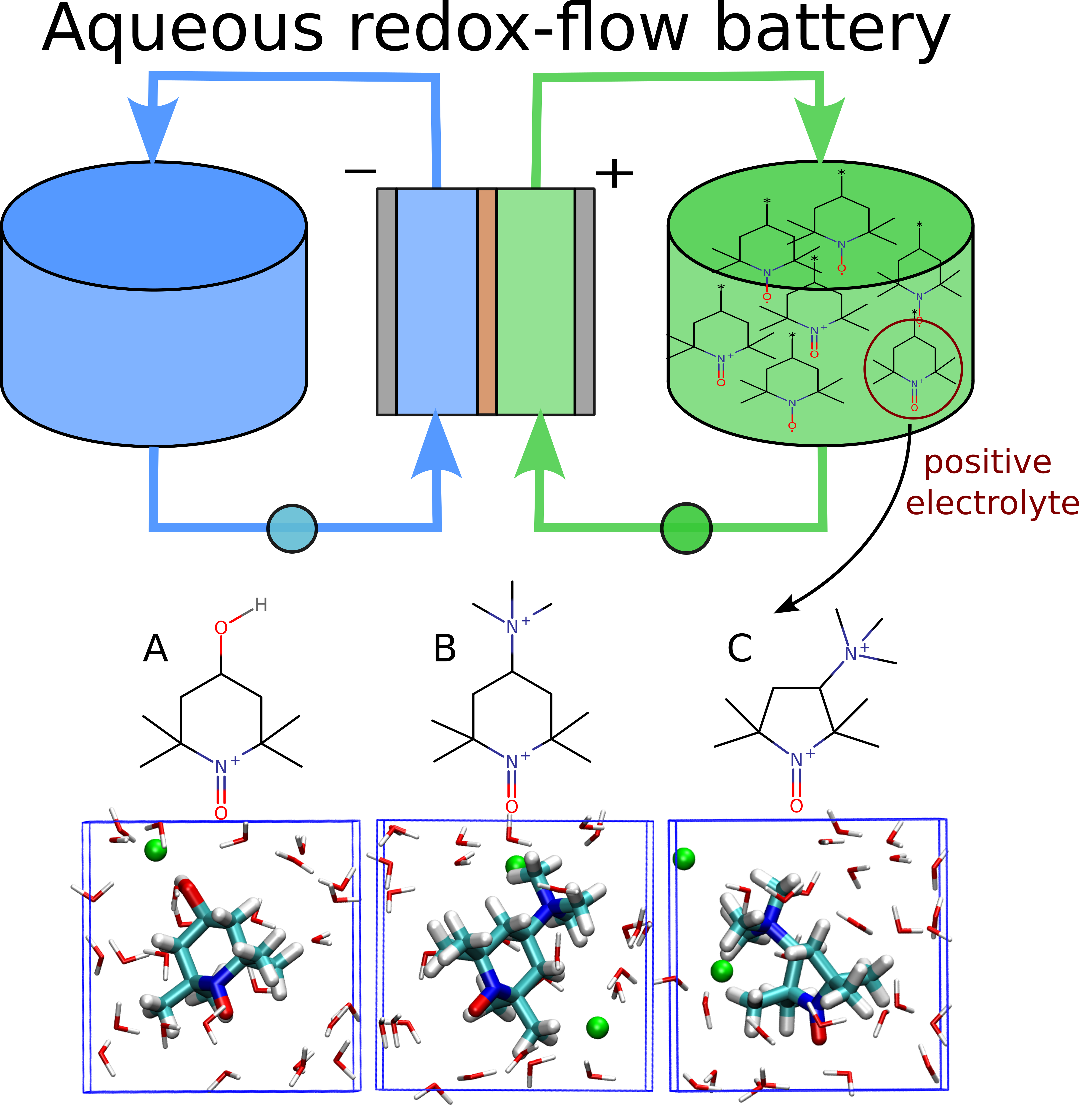}
    \caption{Overview of the systems investigated in this work. TEMPO derivatives are considered as redox-active species in the positive electrolyte of an aqueous redox-flow battery. The three derivatives studied are (A) 4-OH-TEMPO or TEMPOL, (B) 4-TMA-TEMPO, (C) 3-TMA-PROXYL. Representative snapshots of the corresponding solvated simulation box are shown below the structures. Cl$^-$ counter-ion are shown in green}
    \label{fig:systems}
\end{figure}

In this letter, we combine AIMD with enhanced sampling techniques to probe the stability of three TEMPO derivatives in the \(N\)-oxoammonium state and elucidate their degradation mechanisms. To this end, extensive simulations are carried out for each of the proposed degradation pathways. Free energy landscapes are obtained via converged umbrella sampling simulations and orbital energies are computed using electronic structure techniques. Lastly, correlations between orbital and free energies are examined to assess whether orbital energies can serve as a computationally efficient proxy for molecular stability.

\section{Results and discussion}
When several competing degradation mechanisms are possible, their relative likelihood can be assessed by comparing the corresponding free energy landscapes. In particular, the activation free energy provides a direct measure of the barrier associated with each pathway, with lower barriers indicating kinetically more accessible mechanisms. In molecular simulations, these different free energy states are often described using a reaction coordinate , which is a function   mapping the highly-dimensional atomic coordinates \textbf{R} onto a scalar quantity, $s$(\textbf{R}). The free energy along this coordinate is related to its equilibrium probability distribution, $P(s)$, through
\begin{equation}
    F(s)=-k_{\mathrm B}T\ln P(s)+C,
\end{equation}

where $k_{\mathrm B}$ is the Boltzmann constant, $T$ is the temperature, and $C$ is the total Helmholtz free energy of the system. The free energy difference between two states (A) and (B) can then be written as $
\Delta F_{A\rightarrow B}=F(s_B)-F(s_A) $

In practice, however, direct evaluation of the equilibrium probability distribution $P(s)$ from conventional AIMD trajectories is generally not feasible for degradation reactions. Due to computational cost, the accessible simulation timescales are limited to at most tens to hundreds of picoseconds, whereas bond-breaking and bond-forming are rare events and involve free-energy barriers that cannot be overcome through thermal fluctuations. Consequently, the system remains trapped for long periods in the reactant state, and the resulting trajectories do not sample enough of the configuration space to be considered ergodic. The sampled distribution therefore does not provide a statistically meaningful representation of the equilibrium probability, preventing a reliable reconstruction of the free-energy landscape from unbiased AIMD alone. To overcome this limitation, enhanced sampling methods are required to promote transitions along the relevant reaction coordinates and recover the otherwise poorly sampled regions of configuration space. Here, we employ metadynamics simulations~\cite{laio2002escaping} to explore the proposed degradation pathways, or in other words, to obtain reactive trajectories connecting the reactant state to a product state. The main advantage of this simulation setup is that it drives the system to react without necessarily constraining it to reach a predefined product state, allowing alternative products to emerge. Once a reactive pathway is obtained, the corresponding free-energy profile is computed using umbrella sampling~\cite{torrie1977nonphysical}.\\

\begin{figure}[H]
    \centering
    \includegraphics[width=0.9\linewidth]{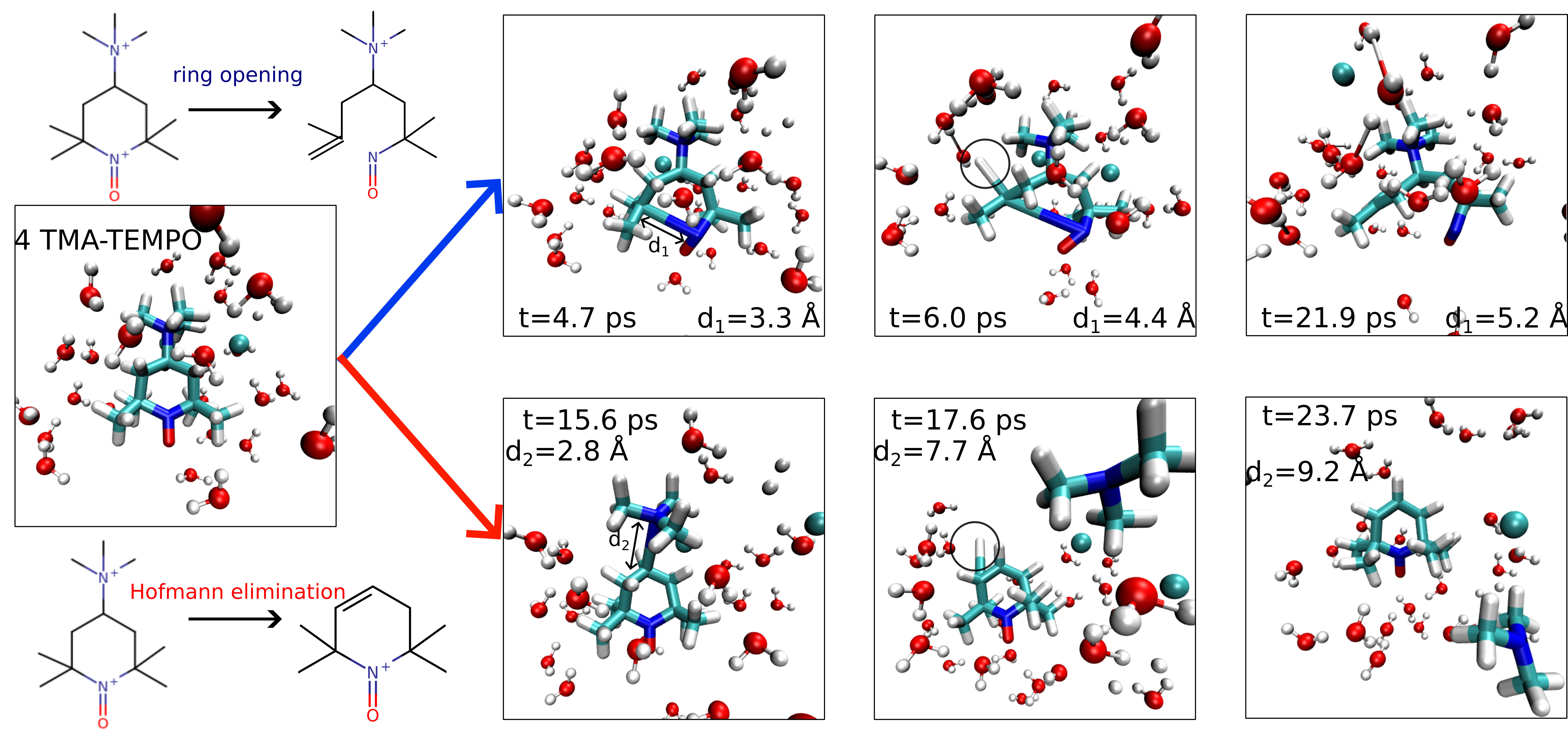}
    \caption{Representative snapshots from metadynamics trajectories of 4-TMA-TEMPO undergoing the two degradation pathways considered in this work. The initial solvated structure is shown on the left. The upper trajectory shows ring opening, obtained by applying a bias along the reaction coordinate $d_1$, while the lower trajectory shows Hofmann elimination obtained by biasing $d_2$.  $d_1$ and $d_2$ correspond to the different C--N distances indicated by the black arrows. Selected configurations are shown at different simulation times together with the instantaneous value of the corresponding reaction coordinate. The circled H atom marks the proton transferred to water during formation of the C=C double bond}
    \label{fig:metadynamics}
\end{figure}

\begin{figure}[H]
    \centering
    \includegraphics[width=0.9\linewidth]{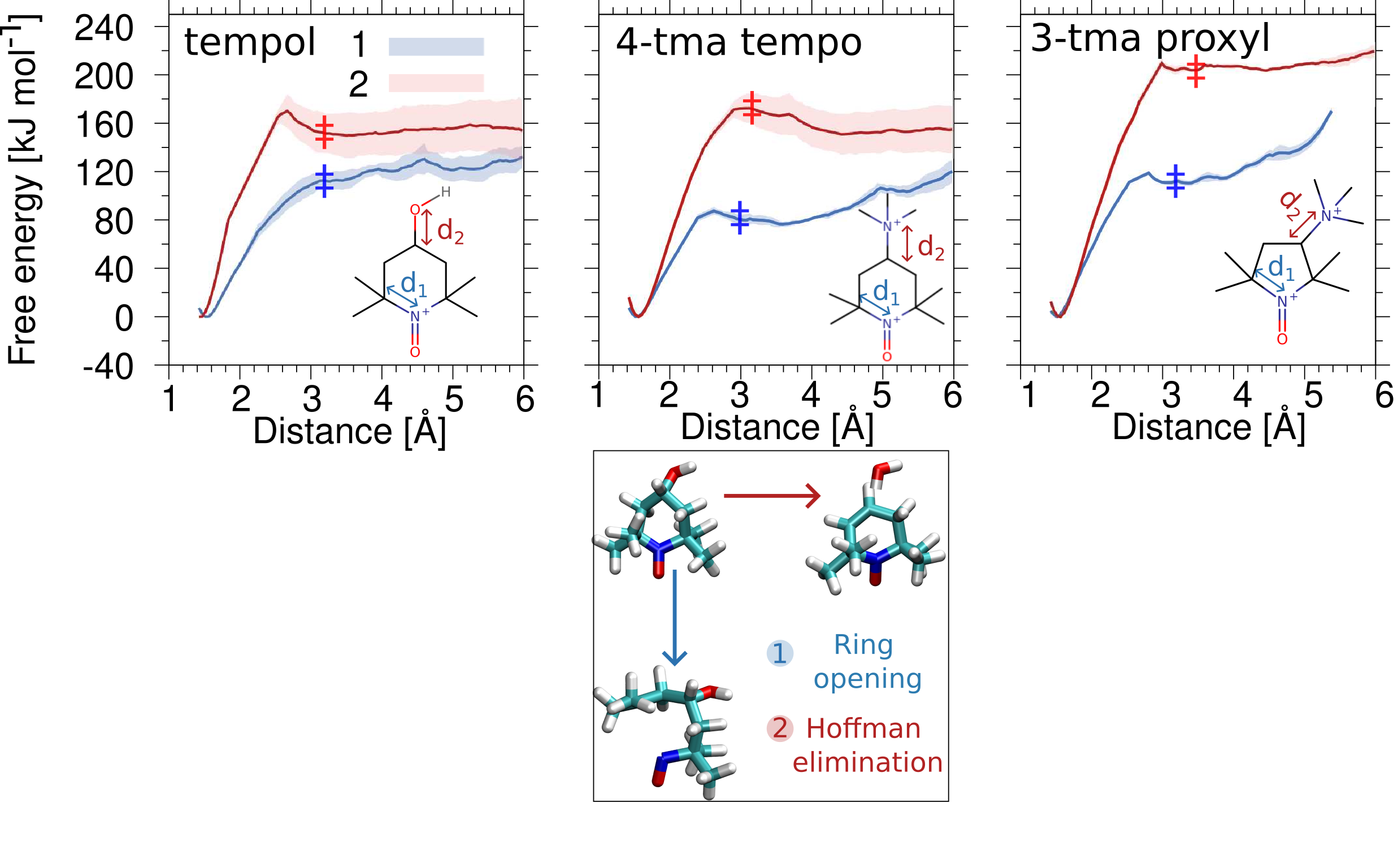}
    \caption{Free-energy profiles obtained from umbrella sampling for the ring-opening (blue, 1) and Hofmann elimination (red, 2) pathways of the three TEMPO derivatives investigated. The corresponding reaction coordinates, \(d_1\) and \(d_2\), are indicated in the molecular structures shown in each panel and are the same as those used in the metadynamics simulations. Shaded regions represent the statistical uncertainty estimated by block averaging over four independent blocks. Daggers mark the committor transition along each reaction coordinate, separating configurations with \(p_B=0\) on the reactant side from those with \(p_B=1\) on the product side. The bottom-right panel illustrates the two degradation pathways for a TEMPOL molecule}
    \label{fig:umbrella-sampling}
\end{figure}

An appropriate reaction coordinate should distinguish the relevant states while capturing the main structural changes along the process. For the degradation reactions considered here, interatomic distances provide a natural choice, since bond breaking and formation are the main events that define the reaction pathways. For the ring-opening mechanism, the reaction coordinate $d_1$ is defined as the distance between the nitroxide N atom and the neighboring C atom of the ring. For the Hofmann elimination, the reaction coordinate $d_2$ corresponds to the distance between the ring C atom bearing the functional group and the directly bonded atom of that group, i.e. O for molecule A and N for molecules B and C.
By applying a bias along the carefully chosen reaction coordinate, we successfully drove the system from the intact oxoammonium state toward degraded products, effectively generating reactive trajectories for each investigated mechanism. In fact, the solute's net charge changes from $+2$ to $+1$ upon degradation hindering its electrochemical potential, as its nitroxide moiety loses its positive charge. Figure \ref{fig:metadynamics} illustrates two such reactive trajectories for the 4-TMA-TEMPO derivative, showcasing the two primary degradation pathways considered in this work: ring-opening degradation and Hofmann elimination. \\ 
Interestingly, the two degradation processes occur over noticeably different timescales in the metadynamics trajectories. In metadynamics, Gaussian bias potentials are progressively added along the reaction coordinate to prevent the system from repeatedly sampling the reactant state and to facilitate barrier crossing~\cite{henin2022enhanced,laio2002escaping}. This observation is particularly meaningful given that a higher Gaussian height, and therefore a more aggressive bias, was required to induce the Hofmann elimination compared with the ring-opening pathway. This suggests that the two mechanisms may possess different kinetic barriers. To quantify these barriers, umbrella sampling was then employed. In this approach, a series of restrained simulations are performed at successive values of the reaction coordinate and combined to reconstruct the free-energy profile. Because the relevant regions of the reaction coordinate are sampled directly, umbrella sampling provides a more efficient way to determine activation barriers than extracting them from the metadynamics trajectories. The resulting free-energy profiles are shown in Figure~\ref{fig:umbrella-sampling}.\\
For all three TEMPO derivatives, the ring-opening pathway presents a lower activation free energy than the Hofmann elimination, indicating that ring opening is kinetically favored under the conditions considered here. The difference between the two pathways is pronounced and the ring-opening mechanism remains the preferred degradation pathway for all molecules investigated. These results confirm the qualitative behavior observed in the metadynamics trajectories. \\ 
To assess the calculated activation barriers, an independent \textit{a posteriori} committor test was performed. In simple terms, the committor, $p_B$, measures whether a given configuration is more likely to evolve toward the reactant ($p_B=0$) or toward the product state ($p_B=1$). The transition state was identified as the configuration for which $p_B \approx 0.5$, corresponding to an approximately equal probability of evolving toward either state. The activation free energy was then taken as the free-energy difference between the reactant minimum and this configuration, indicated by the daggers in Figure~\ref{fig:umbrella-sampling}. The magnitude of the activation barriers varies significantly across the different derivatives. For the ring-opening pathway, the molecules are ranked from lowest to highest activation barrier as 4-TMA-TEMPO, TEMPOL, and 3-TMA-PROXYL.

\paragraph{Discussion.} The favorable ring-opening pathway predicted by the present free-energy calculations is consistent with an increasing body of experimental evidence for the degradation of TEMPO-based electrolytes. Earlier work on 4-TMA-TEMPO proposed a ring-opening mechanism based on gas chromatography and mass spectrometry analyses, although in this case the pathway was inferred from volatile decomposition products rather than directly identified~\cite{nolte2022stability}. More recently, NMR and mass-spectrometry measurements showed that 4-TMA-TEMPO undergoes ring-opening side reactions~\cite{tang2025adjusting}. Most recently, Ref.~\cite{yue2026towards} provided further evidence for ring opening during the degradation of TEMPOL. NMR revealed the formation of alkene functionalities, while GC-MS detected acetone, both consistent with the proposed ring-opening pathway. Although the primary ring-opened species was not directly detected by mass spectrometry, these measurements provide strong experimental support for the mechanism. These studies also illustrate the difficulty of identifying degradation mechanisms experimentally, since short-lived intermediates may undergo further reactions before they can be detected. Thus, the present simulations provide a complementary atomistic picture by directly following the bond-breaking and bond-forming events leading from the oxoammonium reactant to the degraded products. Experimental degradation is, however,  not observed for all derivatives considered here. For 3-TMA-PROXYL, NMR and UV-visible measurements showed little irreversible degradation even after cycling or prolonged storage in the oxidized state~\cite{boutamine20253}. This is still consistent with our results, since ring opening can be the preferred pathway while remaining too slow to occur on the experimental timescale if its absolute barrier is sufficiently high.\\
For the ring-opening pathway, the comparison between structurally related derivatives highlights the effect of both ring size and steric protection. Going from 4-TMA-TEMPO to 3-TMA-PROXYL, while keeping the same trimethylammonium functional group, increases the activation barrier by about 29~kJ~mol$^{-1}$, showing that the five-membered ring is less prone to ring opening than the six-membered TEMPO ring. 

\section{Computational methods}
Three tailored molecules of the nitroxides family are considered: A) 4‑hydroxy-TEMPO; B) 4-TMA-TEMPO (the state-of-the-art compound proposed in Ref.~\cite{janoschka2016aqueous}); C) 3-TMA-PROXYL (a compound with similar redox potentials as (B) proposed in Ref~\cite{boutamine20253}. For each system, the nitroxyde is surrounded by 27 water molecules to ensure a concentration of $2.1$~M. Two counter-ions Cl$^{-}$ are added for systems B and C and one Cl$^{-}$ for A to have an electrically-neutral simulation box. The molecular structures of the three systems can be found in Fig. \ref{fig:systems}. All initial molecular geometries are obtained via geometry optimization using ORCA~\cite{neese2012orca,neese2020orca} at the PW91/TZVP level with implicit solvation (CPCM, water). Initial simulation boxes are equilibrated through short classical molecular dynamics simulations using the AMBER99SB-ILDN force field~\cite{lindorff2010improved} and the TIP3P water model~\cite{jorgensen1983comparison} as implemented in the GROMACS~\cite{abraham2015gromacs} software package.

All MD simulations are performed within the Born-Oppenheimer framework using the CP2K~\cite{kuhne2020cp2k} software package. The electronic structure is treated at the GGA level using the PBE~\cite{perdew1996generalized} exchange-correlation functional, supplemented with Grimme's D3 dispersion correction~\cite{grimme2010consistent} to account for van der Waals interactions. The GPW scheme as implemented in the QUICKSTEP module is employed to represent the electronic density using a dual basis of Gaussian orbitals and plane waves, with a cutoff of 400 Ry. Valence electrons are described using the DZVP-MOLOPT-SR-GTH basis set together with GTH pseudopotentials~\cite{goedecker1996separable,hartwigsen1998relativistic,vandevondele2007gaussian}. The Kohn-Sham equations are solved self-consistently using the orbital transformation method with a convergence threshold of $10^{-6}$. Trajectories are propagated in the NVT ensemble at 330 K with a timestep of 0.5 fs, using a CSVR thermostat~\cite{bussi2007canonical} with a time constant of 100 fs.

For the biased simulations, we employ a two-step approach. In the first step, well-tempered metadynamics~\cite{barducci2008well} as implemented in PLUMED~\cite{tribello2014plumed} is used to explore the degradation pathways of the \(N\)-oxoammonium form. For each of the three systems, two one-dimensional metadynamics simulations are performed, each biased along a distinct interatomic distance as the reaction coordinate, corresponding to the possible degradation mechanisms. For instance, ring-opening pathways are investigated by biasing the distance between the nitrogen atom of the N=O moiety and its neighboring carbon atom, progressively stretching the bond until cleavage occurs. Since some pathways are inherently more accessible than others, the bias factor and Gaussian height are tuned individually for each system and pathway. Throughout these simulations, the system remains free to explore the configurational space along the chosen reaction coordinate, allowing spontaneous identification of intermediates and transition states that would otherwise be inaccessible on AIMD timescales. In the second step, umbrella sampling~\cite{torrie1977nonphysical} is performed along the same CVs to obtain quantitative free energy profiles. Approximately 20 windows are distributed across each reaction coordinate, with stronger force constants applied in the reactant state region to ensure sufficient sampling. The resulting biased distributions are combined using the weighted histogram analysis method (WHAM)~\cite{kumar1992weighted} to reconstruct the free energy profiles. \\
To further characterize the transition region along each degradation pathway, an \textit{a posteriori} committor analysis was performed on configurations extracted from the reactive trajectories. For each selected configuration, 50 short unbiased AIMD trajectories of approximately 0.5 ps were initiated and classified according to whether they evolved toward the reactant or product basin. The committor probability, $p_B$, was then defined as the fraction of trajectories reaching the product state. Approximately 30 configurations were analyzed for each degradation mechanism and molecule. The committor analysis was used to locate the transition between configurations returning to the reactant basin $p_B=0$ and those evolving toward the product basin $p_B=1$; this position was then used to identify the activation free energy on the umbrella-sampling profile.
%Full details of the metadynamics and US set-up are provided in the Supporting Information.

\section*{Acknowledgements}
This work was supported by the France 2030 program, project RADICAL (Grant ANR-23-PEBA-0005). The authors
acknowledge HPC resources granted by GENCI, France (resources of CINES, Grant No.A0190910463). The authors thank Olivier Ouari, Steven Le Vot, and Mathieu Etienne for helpful discussions concerning the experimental degradation mechanisms. L.M. thanks Arthur France-Lanord for guidance in setting up the umbrella sampling simulations. 

%%%%%%%%%%%%%%%%%%%%%%%%%%%%%%%%%%%%%%%%%%%%%%%%%%%%%%%%%%%%%%%%%%%%%
%% If you are using classical BibTeX rather than biblatex,
%% remove the \printbibliography and uncomment the \bibliograpy one
%%%%%%%%%%%%%%%%%%%%%%%%%%%%%%%%%%%%%%%%%%%%%%%%%%%%%%%%%%%%%%%%%%%%%
\printbibliography
%\bibliography{acs-template.bib}

\end{document}